# Why not Neutrinos as the Dark Matter? A Critical Review, KATRIN and New Research Directions

D.J. Buettner  
Independent Researcher  
Dr.Doug_B@yahoo.com

P.D. Morley  
Blue Ridge Scientific, LLC  
peter3@uchicago.edu

**Abstract**

We challenge the traditional wisdom that cosmological (big bang relic) neutrinos can only be hot Dark Matter. We provide a critical review of the concepts, derivations and arguments in foundational books and recent publications that led respected researchers to proclaim that "[Dark Matter] cannot be neutrinos". We then provide the physics resulting in relic neutrino's significant power loss from the interaction of its anomalous magnetic moment with a high-intensity primordial magnetic fields, resulting in subsequent condensation into Condensed Neutrino Objects (CNOs). Finally, the experimental degenerate mass bounds that would rule out condensed cosmological neutrinos as the Dark Matter (unless there is new physics that would require a modification to the CNO Equation of State) are provided. We conclude with a discussion on new directions for research.

**Introduction**

The astronomical and astrophysics community eventually concluded that "crazy Fritz" [1] Zwicky was correct, i.e., that there was in fact a '*dunkle (kalte) materie*' affecting the center-of-mass motion of constituent galaxies in galaxy clusters [2]. For example, the Coma Cluster has fastest member galaxy ~ 3100 km/s relative to the Coma center-of-mass [3].[1] This large relative speed indicates that these "Dark Matter Objects" (DMO) have masses on the order of ~$10^{15}$ solar masses and radii on the order of ~2000 kiloparsecs (kpc). The investigation of what causes these extremely large virial velocities and "flat rotation curves" has resulted in a multi-decade search for what Dark Matter is [4, 5]. The list of hypothesized causes for the Dark Matter phenomenon is quite extensive (see Table 1). Yet, to date no one has been able to experimentally confirm any of them as the Dark Matter [4, 5, 6, 7, 8, 9, 10, 11, 12].

**Table 1. Search Status for Dark Matter Searches**

| Category | Hypothesis | Experimental Status |
|---|---|---|
| Weak Scale | Super-symmetric particles | None observed, search continuing [13, 14, 15] |
| | Extra-dimensions | Theoretical, not observed [16, 17, 18, 19] |
| | Little Higgs | None observed [20] |
| | Effective Field Theory | None observed yet, detectors under consideration/development [21, 22] |
| | Simplified Models | None observed, theoretical [23, 24] |
| Modified Gravity | MOG | Mostly ruled out [25, 26, 27] |
| | MoND | Mostly ruled out [25, 26, 27] |
| | TeVeS | Ruled out [25, 26, 27] |
| | Emergent Gravity | Mostly ruled out [25, 26, 27] |
| Light Bosons | Fuzzy Dark Matter | Partially ruled out [28, 29] |
| | QCD Axions | Not observed, partially ruled out searches continue [30, 31, 32, 33, 34, 35, 36] |
| | Axion-like Particles | Not observed, partially ruled out searches continue [29, 30, 31, 32, 33, 34, 35] |
| Macroscopic | WIMPs | "Almost" completely ruled out [37, 38, 39] |

---

[1] The reference [3] provides the "line-of-sight" velocity distribution in Figure 5 and a 1σ velocity dispersion of 1038 km/s.

| Category | Hypothesis | Experimental Status |
|---|---|---|
| | Macros | "Almost" ruled out [40, 41, 42, 43, 44, 45] |
| | MaCHOs | Ruled out as a viable DM candidate [46] |
| | Primordial Black Holes | Ruled out as a viable DM candidate [47, 48] |
| Other Particles | WIMPzilla | Partially ruled out, possible NASA POEMMA mission to continue search [49] |
| | Superfluid-Ultra light DM (Bose-Einstein Condensates (BEC), Boson stars, etc) | Not ruled out experimentally [50, 51, 52]* |
| | Self-interacting | Not observed, astrophysical constraints [53, 54, 55] |
| Neutrinos | Standard Model (SM) - Dirac | Ruled out, as a thermodynamic gas [56, 57] |
| | Condensed SM - Dirac | Not ruled out [56, 57]** |
| | SM - Majorana | Theoretical, essentially ruled out but experiments continue [58, 59, 60, 61, 62, 63, 64, 65] |
| | Sterile | Almost ruled out, but searches continue [66, 67, 68, 69, 70, 71, 72, 73, 74] |
| * We argue that a BEC and Boson stars, following Bose-Einstien statistics lacks a mechanism to prevent gravitational collapse into a black hole at the masses required to explain gravitationally bound Dark Matter objects that are on the order of ~$10^{15}$ $M_O$) consistent with galaxy clusters. ||| 
| ** Traditional arguments used to claim that neutrinos cannot be Dark Matter only apply to neutrinos as a thermodynamic gas. This paper provides the arguments that condensed neutrinos are the Dark Matter. |||

Additional reviews, and press releases are also found in references [75, 76, 77, 78, 79, 80, 81, 82]. A good summary of the entire approach to identify Dark Matter was summarized by Cushman in [39] as,

"Every time that we get to a higher sensitivity and don't see something, we have made some very definitive statements about the nature of dark matter," Cushman says. "They're negative statements, if you like, but they are absolutely changing the way we look at the world. And without those negative statements, we would not try for another model. We would not say that we have to work harder for something different than what we thought it might be."

In the next section, we critique the arguments from influential books which identifies why numerous researchers and authors have all but abandoned cosmological (relic big bang) neutrinos as the early front runner to be the "Dark Matter", even to the point of proclaiming that "[Dark Matter] cannot be neutrinos" [6, 7, 83]. Afterwards, we describe the physics leading to neutrinos radiating away their power in the presence of chaotic primordial magnetic fields such that they never were in thermodynamic equilibrium with baryonic matter. We then describe the resultant Condensed Neutrino Objects (CNOs), with masses on the order of ~$10^{15}$ solar masses and radii on the order of ~2000 kiloparsecs (kpc). We demonstrate how CNOs satisfy the local group's rotational velocities, show that the background neutrino density is within limits set by KATRIN, and finish with a discussion on the implications of our arguments with recommendation on new experimental directions.

**The Arguments Used to "Rule Out" Cosmological (Relic) Standard Model Neutrinos**

In Table 2 we provide the arguments in various books that these authors use to advance the claim that cosmological (relic) "standard model neutrinos" are not the Dark Matter.

**Table 2. Book Arguments and our Critiques**

| Publication Date | Books/Papers | Summarized Arguments/Assumption | Critique(s) |
|---|---|---|---|
| 1990 | Kolb & Turner [84] | General treatise of different potential neutrino mass ranges, species types (Dirac or Majorana) and numbers (3 varieties with antineutrino and 3+1 sterile variety), does discuss de Sitter phase excitations of other very light fields, and "In the case of the photon field, it is possible that the de Sitter-space-produced fluctuations ultimately lead to the generation of large-scale primeval magnetic fields". [85] | Doesn't recognize a primordial magnetic field induced neutrino power loss and subsequent condensation, treats neutrinos as "hot Dark Matter", and treats "cold Dark Matter" as being made up of WIMPS. |
| 1993 | Peebles [86] | In the section on Dark Matter, good discussion on why "Interest shifted away from massive neutrinos [as the Dark Matter]", (1) experimental bounds on the mass for the electron-dominated family was reduced below $m_\nu =$ 93 $\Omega_\nu h^2$ eV (where $\Omega_\nu$ is the neutrino density parameter) for interesting values of Hubble's constant, (2) recognized that primeval fluctuations are Gaussian, scale-invariant, and adiabatic (dominant mass components all have the same primeval space distribution), although (1) and (2) were not deal breakers, (3) the most influential argument was the recognition that other nonbaryonic dark mass candidates covered possibilities for galaxy-formation models. He does briefly discuss primordial magnetic fields. [87] | Peebles does briefly discuss primordial magnetic fields but doesn't acknowledge how primordial magnetic fields may induce neutrino power loss and subsequent condensation of neutrinos. |
| 2008 | Weinberg [88] | Describes how L-particles (left-over) was originally thought to be heavy neutrinos, but that "They could not be any of the three known types', which would at most have masses on the order of 1 eV and continues to discuss the possibility of very heavy lepton with negligible mixing but states "that the cold Dark Matter could not consist of a new heavy neutrino". Further, Weinberg contends that the most plausible candidate is one of the new particles required by supersymmetry. [89] | No primordial magnetic field power loss treatment; no condensation treatment; assumes neutrinos only exists as a thermodynamic gas. No evidence of supersymmetric particles. |
| 2021 | Dodelson & Schmidt [90] | Assumes neutrinos are *hot* with large velocities, thus unable to clump in the early universe, subsequently treats the density parameters for Dark Matter and neutrinos as distinctly separate. [91] | No cosmological magnetic field treatment to allow for power loss and subsequent rapid cooling. |

Peebles [87], stated that originally massive neutrinos were considered as a viable Dark Matter candidate, but when the experimental mass of the neutrino fell below what was considered as feasible (from a thermodynamic gas standpoint), the belief was that they could only be "hot Dark Matter" in the early universe; at this point the community essentially abandoned neutrinos in favor of other particle types. Our primary critique, centers around a lack of realization that the existence of chaotic primordial magnetic fields from the expanding plasma provide a power loss mechanism for neutrinos to significantly cool and condense [92]. **This condensation mechanism was not considered by any of these cosmology book references.**

We will discuss in the next two sections how standard model Dirac neutrinos exhibit power loss in the presence of primordial magnetic fields and then subsequently condense to form Dark Matter Objects.

**Primordial Magnetic Fields (PMF) and the Hubble Tension**

Arguments on the possible existence of primordial magnetic fields (PMF) can be found in Peebles [87] and [99, 100, 101, 102]. Further, even though the Planck consortium published limits on the PMF [103] (with additional references to the effects of PMFs on the CMB can be found in [104, 105] ), their Λ-Cold Dark Matter (Λ-CDM) assumptions result in a 5-sigma discrepancy with other methods [113], suggesting that one or more of the Λ-CDM model assumptions are wrong.

While there are other ways to resolve the issue of the Planck group's baryon acoustic oscillation observations [107, 108, 109, 110], other physicists [111, 112] have independently joined us in suggesting that the Hubble tension [113] could be resolved using PMFs. We contend that the Λ-CDM model's lack of appropriately accounting for the impact of primordial magnetic fields on the Cosmic Microwave Radiation (CMR) signature and the neutrino power loss mechanism is the fundamental reason for the Hubble tension. Replacing the baryonic acoustic degrees of freedom provided by thermodynamic equilibrium neutrinos with magnetic field degrees of freedom should resolve the Planck issue.

We will call the revised Cold Dark Matter model the νCDM (pronounced nu-CDM) to deliberately point out that a chaotic PMF resolution to the Hubble tension and its interaction with the neutrino's anomalous magnetic moment [114, 115] also resolved the Dark Matter identification problem. This neutrino – chaotic PMF interaction causes spin-flipping power loss allowing neutrino condensation during big bang expansion.

The result, from the power loss mechanism, which we will explore in the next section, is that neutrinos were never in thermodynamic equilibrium with baryonic matter, and that Dark Matter is in fact degenerate condensed neutrino matter.

**Condensation of Cosmological Neutrinos**

If neutrinos are Dirac particles, they will have tiny magnetic moments. However, since Early Universe magnetic fields are expected to be large in value and chaotic in nature [114], they give rise to magnetic cooling [92],

$$P_{cooling} = \frac{2}{3c^3} \left[ \gamma^8 \left( \vec{B} \cdot \dot{\vec{B}} \right)^2 \dot{\vec{\mu}}^2 + 2\gamma^6 \left( \vec{B} \cdot \dot{\vec{B}} \right) \left( \dot{\vec{\mu}} \cdot \ddot{\vec{\mu}} \right) + \cdots \right]. \quad (1)$$

The leading term $\left( \gamma = \sqrt{1 - \vec{\beta}^2}, \vec{\beta} = \vec{v}/c \right)$ in Eqn. 1 for the radiation loss ($P_{cooling}$) does not depend on the tiny magnetic moment $\vec{\mu}$, but instead on its magnitude of its flipping $\dot{\vec{\mu}}^2$, which ranges from 0 to ∞. Electrons, on the other hand carry electric charge and gain energy through Early Universe electric fields. This magnetic cooling mechanism, in conjunction with the space-time expansion, can lead to neutrino condensation.

**Condensed Neutrino Object Physics**

The Local Group of Galaxies seems to be embedded in a Dark Matter Object (DMO) [99]. We describe in these sections how a Condensed Neutrino Object (CNO), which are stable assemblages of neutrinos and anti-neutrinos, are a candidate for the DMO in the local group of galaxies [121]. The following is provided from reference [119].

Let us discuss the situation of a spiral galaxy embedded in a Coma-like Galaxy Cluster Dark Matter potential and see the complexity of the resulting gravitational potential. The Dark Matter gravitational potential at position (**r**) inside the dark object is

$$\Phi(\mathbf{r}) = -G \int \frac{\rho_B(\mathbf{r}')}{|\mathbf{r} - \mathbf{r}'|} d^3\mathbf{r}', \qquad (2)$$

G is the gravitational constant, where $\rho_B$ is the Dark Matter mass density. For Coma or local group-like Dark Matter object, we assume that the density is approximately spherically symmetric[2]; $\rho_B(\mathbf{r}') = \rho_B(r')$ so Eqn. (2) becomes

$$\Phi(\mathbf{r}_c + \mathbf{a}) = -G \int_0^{|\mathbf{r}_c + \mathbf{a}|} \frac{dM_B(\mathbf{r}')}{r'} \qquad (3)$$

where $dM_B(\mathbf{r}') = \rho_B(\mathbf{r}') 4\pi r'^2 dr'$, $\mathbf{r}_c$ is the radius from the origin of the DMO to the spiral galaxy center of mass, and $\mathbf{a}$ is the spiral arm vector. The embedded spiral galaxy is shown in Figure 1 with the vectors from the center of the DMO to the center of the galaxy and to a location in the plane of rotation based on the relative tilt (tilt is the angle between the galaxy's rotation vector and radius to the DMO) with respect to the DMO's center. A graph embedded in the figure depicts the notional relative minimum and maximum speeds of stars in spiral arms is based on the location from the center of the spiral galaxy's center and spin vector orientation. It should be noted that our derivation is a static "snapshot" and ignores the dynamic interaction effects of galaxies in a DMO.

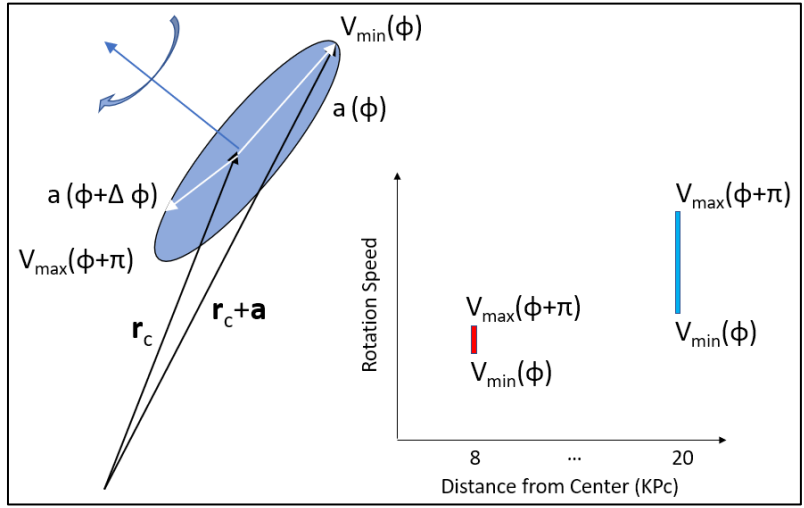

**Figure 1. DMO and Spiral Galaxy Vector Geometry with Relative Rotational Speeds, with the effect of the tilted rotation seen in Figure 3b.**

---

[2] We have documented the potential "breathing modes" [117] of CNO and currently neglect the potential for colliding CNO or nearby CNO gravitational effects.

We're interested in the difference of the gravitational potential between a spiral arm and its galaxy's center of mass, $\Phi(a) = \Phi(\mathbf{r}_c + \mathbf{a}) - \Phi(\mathbf{r}_c)$, as shown in Eqn. (4),

$$\Phi(\boldsymbol{a}) = -G \int_{|\mathbf{r}_c|}^{|\mathbf{r}_c+a|} \frac{dM_B(\mathbf{r}')}{\mathbf{r}'} \tag{4}$$

Consider now circumnavigation 360 degrees around the galaxy center from a fixed distance. This angle becomes the spiral galaxy's azimuthal angle ϕ. If the spin axis is tilted with respect to a radial, then Eqn. (4) has both positive and negative values: for some ϕ: $|\mathbf{r}_c + \mathbf{a}| > |\mathbf{r}_c|$ and for 180 degrees further in ϕ: $|\mathbf{r}_c + \mathbf{a}| < |\mathbf{r}_c|$. Circum-location means a rotating star with a fixed distance from the spiral galaxy's center will go up a potential hill for half its revolution and lose rotational speed. During the other half of the rotation, it will go down a potential hill and gain speed. Thus, the spiral arm rotational speeds are azimuthal angle dependent in Eqn. (4). If however, the spin axis is parallel or antiparallel to a Dark Matter Object radial, then $|\mathbf{r}_c + \boldsymbol{a}|$ = constant > $|\mathbf{r}_c|$, and rotation speeds (for constant distance from center) are no longer azimuthally dependent on the DMO gravitational potential. In the interesting case that $\boldsymbol{a} \to 0$, in very short distant scales (~1 Kpc), then there is almost no change in the DMO potential ($\Phi(a) \to 0$) resulting in star speeds that are nearly the same near each other.

This geometric dependence of rotation speeds always occurs in tilted spiral galaxies embedded in CNO and leads to our claim of the erroneous perception of rotation speeds being attributed to a galactic Dark Matter "halo" [94].

**Local Group in a CNO**

The CNO Equations of State Eq.s (5, 6) (derived in reference [121]) are,

$$\rho_{total} = \frac{6\pi m_\nu^4 c^3}{h^3} \left\{ 2x(1+x^2)^{\frac{3}{2}} - x\sqrt{1+x^2} - \sinh^{-1} x \right\}, \tag{5}$$

$$P_{total} = \frac{2\pi m_\nu^4 c^5}{h^3} \left\{ x(2x^2 - 3)\sqrt{1+x^2} + 3\sinh^{-1} x \right\}, \tag{6}$$

It was shown in reference [121] that the Local Group is embedded in a CNO based on the rotational speeds of the Milky Way, M31 and M33 spiral galaxies. That analysis is considered preliminary as the neutrino mass scale has not yet been experimentally determined. Our recent work indicates that the Local Group DMO is a CNO with x(0) boundary conditions that lie somewhere in between 0.014 and 0.029. The variable x is

$$x \equiv {p}/{m_\nu c}, \text{ and } x(0) = {p_f}/{m_\nu c}, \tag{7}$$

with $p$ being the momentum, and $p_f$ being the Fermi momentum. x is called the reduced momentum and is the momentum ratio. $m_\nu$ is the degenerate neutrino mass scale measured by KATRIN, and c is the speed of light. The assumption is that the CNO have approximately equal numbers of neutrinos and anti-neutrinos. For purposes of the equation of state, neutrinos form a perfect flavor symmetry. The final statistical description is equilibrium with a single Fermi momentum. The value of x(0) = 0.02 means the Fermi momentum has speed = 0.020c. This is the largest value of x, at the center of the CNO. As the radial coordinate goes from the center to the actual radius of the CNO, the momentum ratio also goes to zero (x[r→CNO$_{max\ R}$] → 0) as shown in Figure 2. The figure is found by solving the equation of hydrostatic equilibrium using the condensed neutrino equation of state.

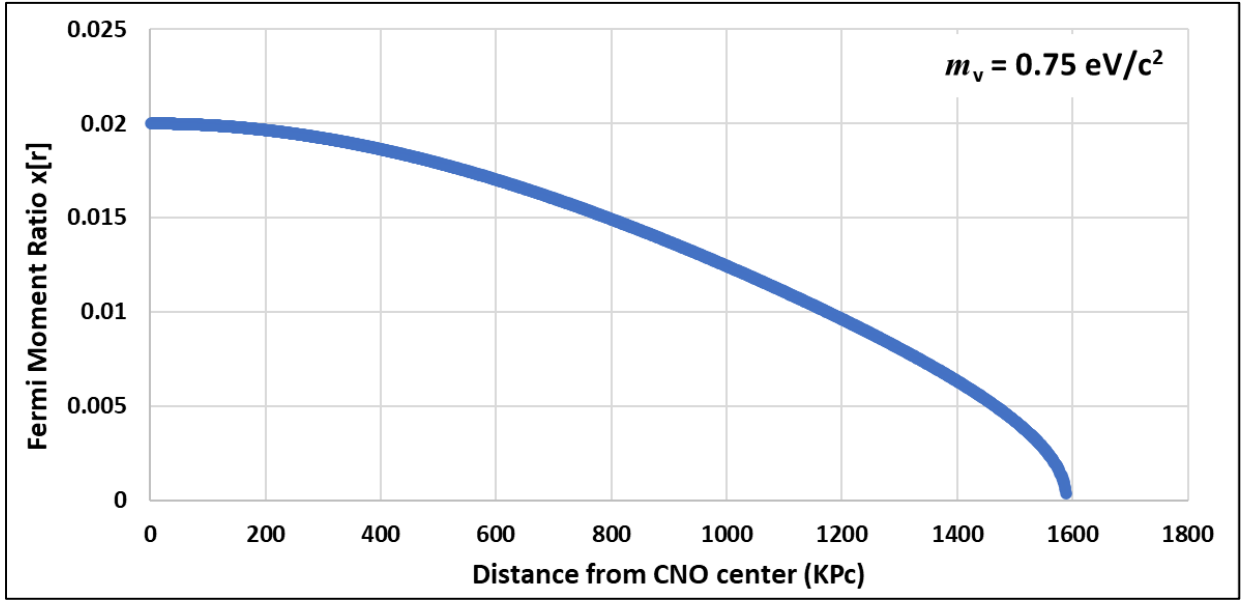

**Figure 2. Plot of the CNO Fermi Momentum ratio x[r] for x[0] = 0.02 with $m_v$ = 0.75 eV/c²**

The Local Group CNO solution we provided in reference [121] assumed a neutrino degenerate mass $m_v$ is ~0.8 eV/c², which currently remains at the upper limit of the KATRIN neutrino degenerate mass. Below we revisit the Local Group to identify plausible bounds on the neutrino's degenerate mass assuming DMO are CNO using the Milky Way's rotational curves and [123] to identify approximate rotation speeds of 8 and 20 kiloparsecs (kpc) as bounds on the Fermi momentum ratio's boundary condition at the center.

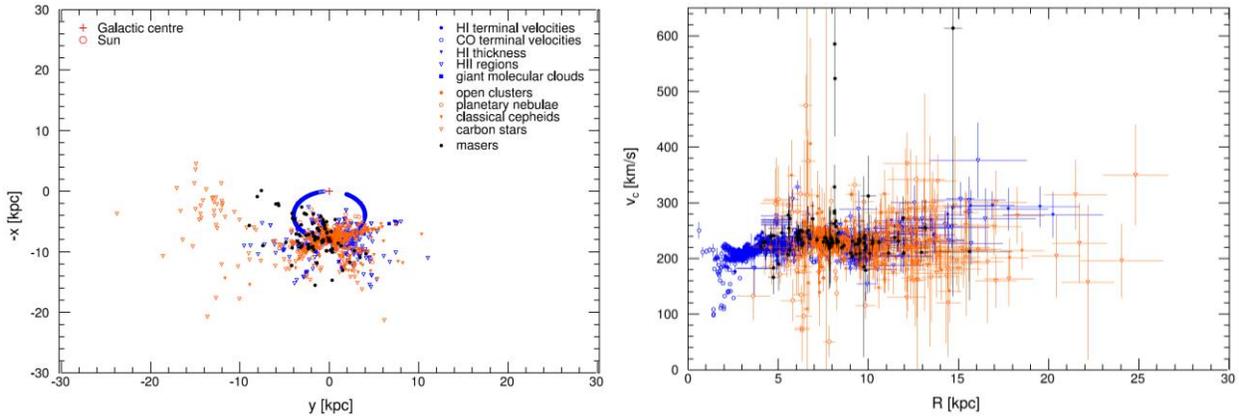

**Figure 3. Left (Fig 3a) is the position tracers in Milky Way with the right (Fig 3b) are the circular velocities of the tracers as derived from gas kinematics (blue), star kinematics (orange) and masers (black) (figures are from figure 2 in [123] reused with permission)**

Compare Figure 3b with the inset Figure 1, where Figure 3b shows that the Milky Way spin axis is tilted with respect to the radial of the DMO. Reference [123] identifies the left panel (the top panel in reference [123]) as depicting the positions of the different tracers in the Galactic plane assuming the Sun's parameters are $R_0$ = 8 kpc, with the galactic center at (x, y) = (0, 0) kpc, and the Sun at (x, y) = (8, 0) kpc. The right panel (the bottom panel in reference [123]) displays the circular velocities of the tracers as a function of the Galactocentric radius assuming $R_0$

= 8 kpc, $v_0$ = 230 km/s and $(U, V, W)_\odot$ = (11.10, 12.24, 7.25) km/s. The authors of reference [123] provide their full data set via their galkin tool on GitHub page github.com/galkintool/galkin which includes data source references used in the figure.[3]

Using the solution from reference [121] for the CNO's location in heliocentric cartesian coordinates (604., -303, 11) kpc and the approach discussed in reference [124] to select appropriate reduced momentum x(0) boundary conditions for a given neutrino degenerate mass, we can approximate the Milky Way's rotation speeds at 8 kpc and 20 kpc from their center as seen in Figure 3. We do the same for M31 (at 30 kpc) and M33 (at 15 kpc) from their centers. An example of the process used for the selection for possible x(0) solutions that bound the rotation speeds for these local group spiral solutions for neutrino degenerate mass of 0.75 eV/$c^2$ is provided in Figure 4.

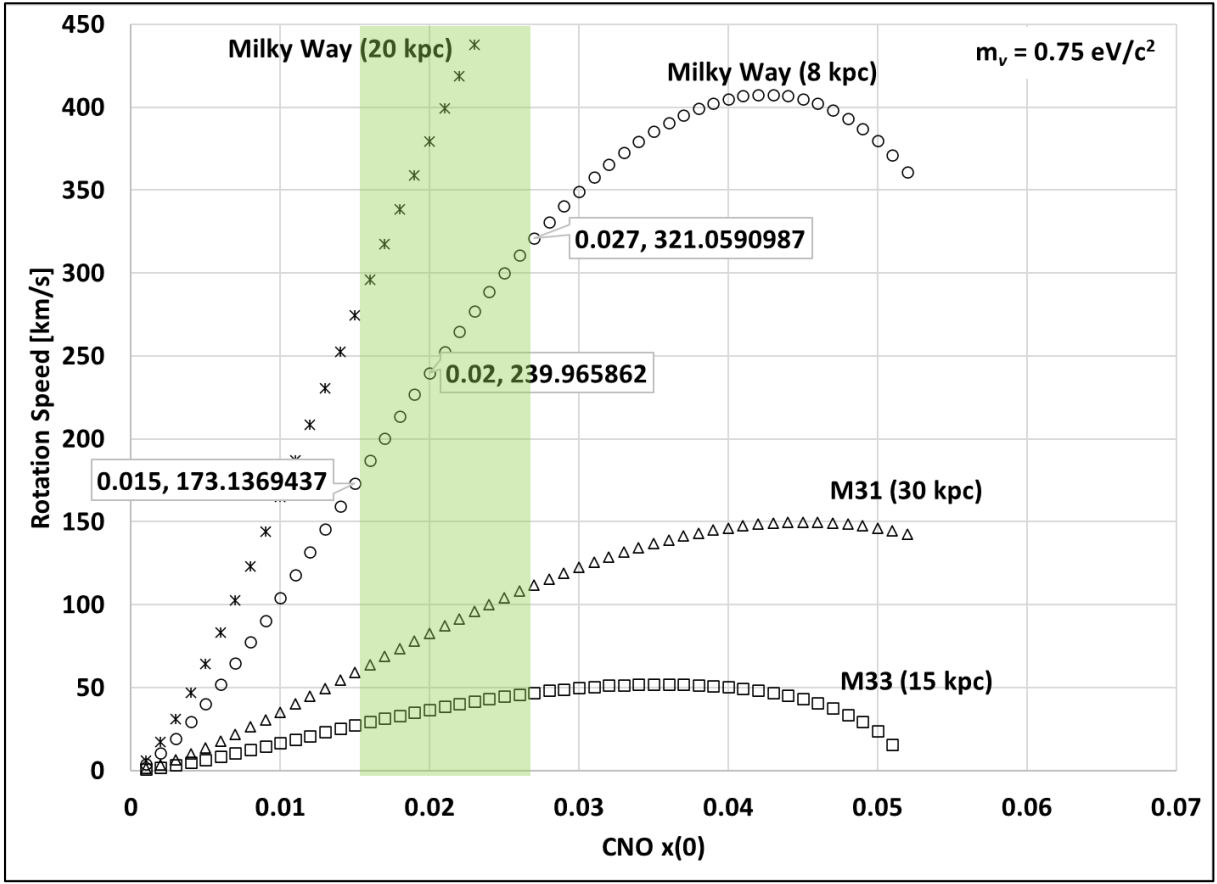

**Figure 4. CNO induced rotation curves for the Milky Way, M31 and M33 for a 0.75 eV/$c^2$ Neutrino Mass**

Following the same approach as used above in Figure 4 for the analysis of other possible degenerate neutrino mass solutions, we identify the viable range of x(0) solutions, and restrict the feasible solution space based on the recent data from KATRIN [56, 57] (degenerate neutrino mass < ~0.8 eV/c$^2$) to plot mass/radii and x[0] bound condition solutions for the local group in Figure 5.

---

[3] We expect the observed significant speed deviations as observed in the Figure 3b to be the result of ancient galactic collisions.

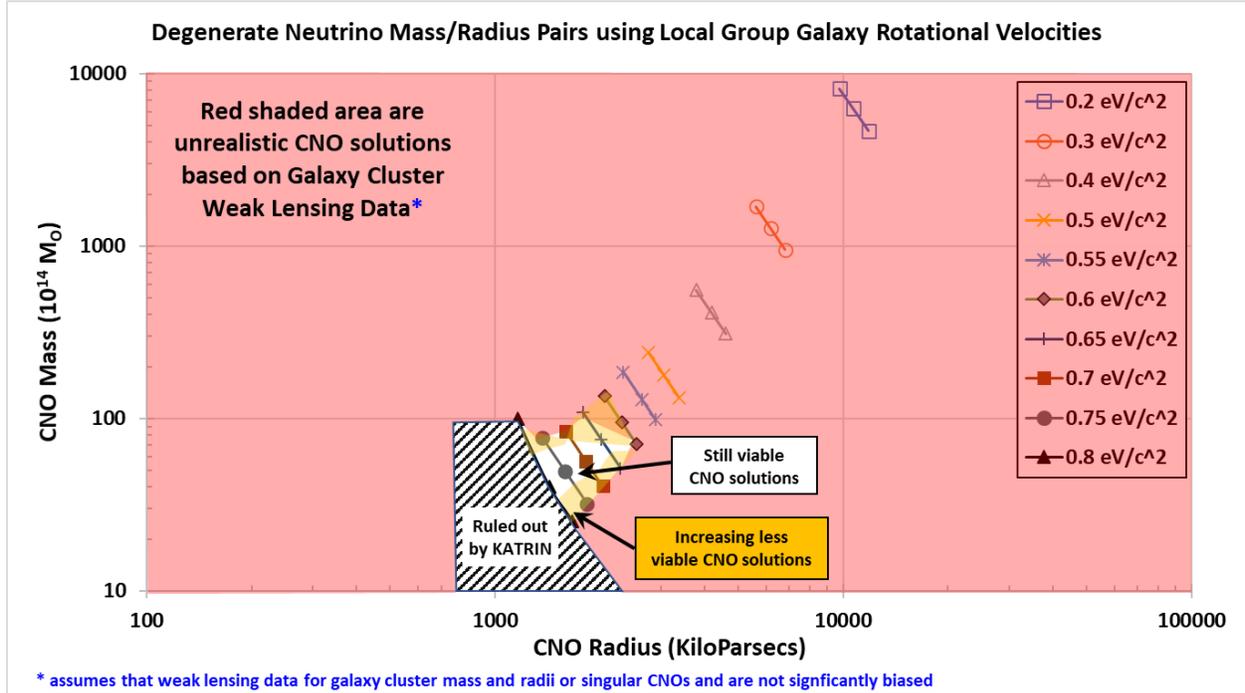

**Figure 5. CNO Mass/Radius Pairs using Local Group Galaxy Dynamics to Show Viable Neutrino Degenerate Mass Ranges**

This figure shows approximate "still viable", increasingly "less viable", and "unrealistic" bounds on possible CNO solutions. We also use weak lensing data analysis from reference [124] to restrict the viable solution space under the assumption that the weak lensing data are not significantly biased in mass or radii as reported by the astronomical community.[4]

In Figure 6, below, we provide an updated 3D location for the CNO encompassing the Local Group of galaxies showing for comparison (Figure 6a on the left) $m_v = 0.75$ eV/c$^2$ and a momentum ratio x(0) boundary condition of 0.02, and (Figure 6b on the right) $m_v = 0.6$ eV/c$^2$ and a momentum ratio x(0) boundary condition of 0.029. Further refinement of the neutrino degenerate mass will allow us to better resolve the orientation of and density boundary conditions for the Local Group CNO.

---

[4] Although, we have argued in the past that CNO do not follow any NFW density profile favored by numerous astronomers, we ignore in this paper the potential biasing effect caused by the use of that density profile.

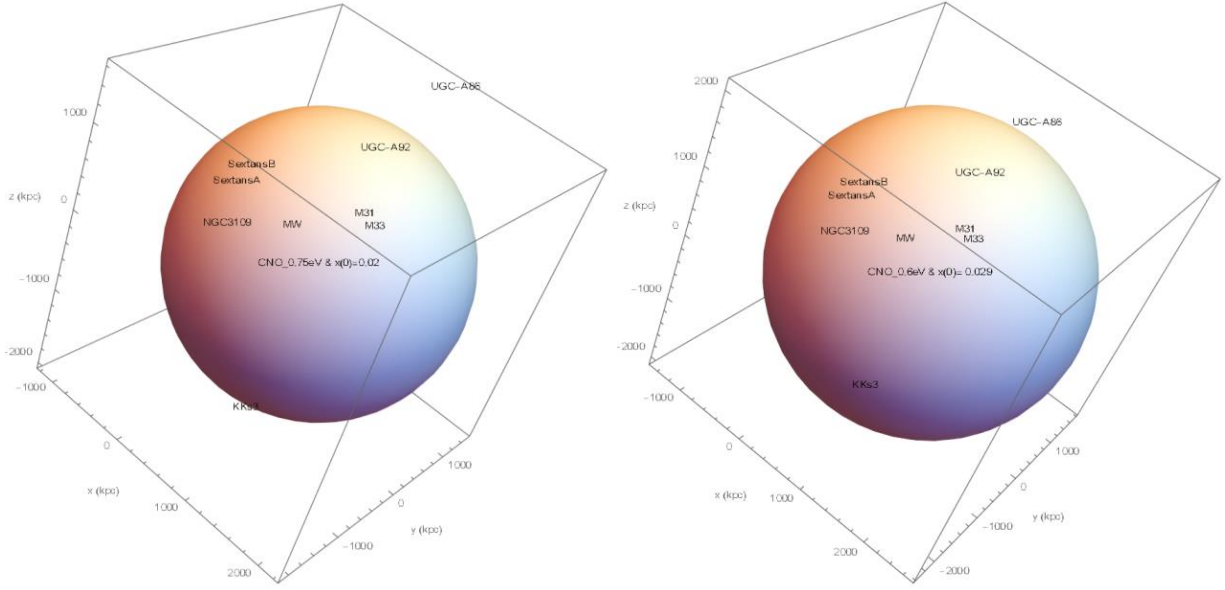

**Figure 6. CNO 3D Location for a Subset of Local Group of Galaxies: (Fig. 6a) on the left with $m_v$ = 0.75 eV/c² and x(0) = 0.02, and (Fig. 6b) on the right with $m_v$ = 0.6 eV/c² and x(0) = 0.029**

The x(0) = 0.02 solution, with $m_v$ = 0.75 eV/c2, has radius ~1.6 Mpc, the estimated distance the Earth is from the center of the CNO is ~675.9 kpc, using Figure 2, gives an estimated x(Earth) ≈ 0.016. We use this result in the next section to calculate the 'overdensity' of neutrino and set a mass correction bound on the KATRIN mass.

**Overdensity of Neutrinos and KATRIN Mass Correction**

KATRIN's tritium is embedded inside a quantum Fermi gas, and the Pauli Exclusion Principle requires that the exiting electron anti-neutrino have energies above this Fermi level. From the estimated location of the Milky Way, we determined in the last section that the top of the quantum gas is around x ≈ 0.016. We now use this to give the KATRIN mass correction: with x ≈ 0.016 means v/c = 0.016. So

$$m_{KATRIN} = m + \frac{mv^2}{2c^2} = m\left(1 + \frac{(0.016)^2}{2}\right) = m(1.000144) \quad (8)$$

The fact that the Milky Way is embedded in a CNO also has implications for the `local relic neutrino density' of 56 electron neutrinos per cm³ [125]. Of course, with the existence of the CNO, there is no relic neutrino density, but a density associated with a CNO. KATRIN [93] has made the measurement that the density ($\eta$) of electron neutrinos < 1.1 x 10¹¹ per cm³ at 90% CI, using the reaction $v_e$ + ³H → ³He + e⁻.

We check that our claimed CNO embedding the Milky Way is below this experimental value. The density of electron neutrinos $N_{ve}$ is

$$N_{ve} = \frac{8\pi}{3}\frac{p_f^3}{h^3} \quad (9)$$

From the previous calculation, $p_f$ = m(0.016c) and m = 0.75 eV/c². Working the numbers reveals $N_{ve}$ ≈ 7.6 x 10⁶, which is significantly less than the 90th percentile KATRIN [93] value.

**Discussion, Recommendations and Conclusions**

In this and our series of papers on Condensed Neutrino Objects (CNO) theory, we have provided not only a mechanism for the condensation of neutrinos early in the universe (magnetic cooling), but we have also demonstrated that they provide a stable massive Dark Matter Object structure with the correct densities to affect the rotation speeds of galaxies depending on their tilt orientation to the center of the CNO.

We continue to propose that if Dark Matter is proven to be condensed neutrinos, then it cannot exist as clumpy "halos", such as described in the quite recent James Webb Space Telescope (JWST) news release [126] which stated "Detecting small halos would be a triumph for the cold dark-matter theory; conversely, not detecting small halos would imply that cold Dark Matter does not exist." We reject this conjecture, as "not detecting halos" is exactly what one should expect with CNO theory. To date we have not yet attempted to derive a condensation rate for CNO due to a lack of sufficient data to properly do so.

Finally, in this paper we demonstrated that CNO theory predicts a neutrino degenerate mass value of ~0.6 to 0.8 $eV/c^2$.

Hence, we expect that if KATRIN has a mass signal, they will announce a neutrino mass range that excludes zero mass, but effectively leaves intact upper limits in the range of ~0.6 to 0.8 $eV/c^2$. Therefore, so long as KATRIN and next generation experiments continue to keep our neutrino mass in their upper bounds, we simply cannot rule out CNO theory as the description of DMOs. Further, we propose that as soon as the KATRIN experiment provides a mass range that rules out < ~0.4 $eV/c^2$ with our upper bound included, we should conclude that the DMO phenomenon is in fact our CNO theory.

Future research that could support the verification that CNOs are in fact the DMOs would be:

- Confirm the power loss mechanism for neutrinos from chaotic magnetic fields
- Dirac vs Majorana neutrino variety experiments
- Redo numerical simulations with cooled/cooling neutrinos from the big bang into CNO and determine if νCDM theory results in a universe that matches observations from astronomy [127, 128, 129]

**Appendix**

Table 3. in this appendix provides the distance from the center of the expected CNO with minimum and maximum galaxy distances in bold font.

**Table 3. Local Group Galaxy Distance from the Center of the Expected CNO**

| Local Group Galaxy | Distance from CNO center (kpc) |
|---|---|
| Antlia 2 | 769.524 |
| Andromeda II | 669.81 |
| Andromeda III | 672.375 |
| Andromeda V | 826.047 |
| Andromeda VI | 524.461 |
| Andromeda VII | 738.873 |
| Antlia Dwarf Galaxy | 1821.67 |

| Local Group Galaxy | Distance from CNO center (kpc) |
|---|---|
| Boötes Dwarf Galaxy | 713.097 |
| Canes Venatici Dwarf Galaxy | 836.474 |
| Canis Major Dwarf Galaxy | 679.967 |
| Carina Dwarf Galaxy | 706.314 |
| **Cetus Dwarf Galaxy** | **394.706** |
| Draco Dwarf Galaxy | 681.038 |
| Fornax Dwarf Galaxy | 628.234 |
| IC 10 | 879.719 |
| IC 1613 | 524.755 |
| KKs 3 | 2028.8 |
| Large Magellanic Cloud | 674.937 |
| Leo A | 1341.06 |
| Leo I | 925.37 |
| M110 | 726.671 |
| M31 | 743.405 |
| M32 | 777.077 |
| M33 | 802.177 |
| Milky Way | 672.864 |
| NGC 147 | 739.616 |
| NGC 185 | 697.44 |
| NGC 3109 | 1832.61 |
| NGC 6822 | 418.095 |
| Pegasus Dwarf Irregular Galaxy | 470.764 |
| Phoenix Dwarf Galaxy | 574.4 |
| Pisces Dwarf Galaxy | 629.804 |
| Sagittarius Dwarf Elliptical Galaxy | 663.84 |
| Sagittarius Dwarf Irregular Galaxy | 718.427 |
| Sculptor Dwarf Galaxy | 621.535 |
| Sextans A | 1995.74 |
| Sextans B | 2017.6 |
| Sextans Dwarf Galaxy | 760.339 |
| Small Magellanic Cloud | 654.954 |
| Tucana Dwarf Galaxy | 742.403 |
| **UGC-A 86** | **2423.23** |
| UGC-A 92 | 1552.93 |
| Ursa Major Dwarf Galaxy | 753.519 |
| Ursa Minor Dwarf Galaxy | 696.42 |
| Willman I | 703.677 |
| Wolf-Lundmark-Melotte | 431.188 |

55. Hayashi, K., et al., "Probing Dark Matter self-interaction with ultrafaint dwarf galaxies", Phys. Rev. D 103, 023017 (2021); doi.org/10.1103/PhysRevD.103.023017
56. The KATRIN Collaboration, "Neutrinos Are Lighter than 0.8 Electron Volts", KATRIN Press Release: https://www.kit.edu/kit/english/pi_2022_012_neutrinos-are-lighter-than-0-8-electron-volts.php
57. The KATRIN Collaboration, "Direct neutrino-mass measurement with sub-electronvolt sensitivity", Nat. Phys. 18 (2022); doi.org/10.1038/s41567-021-01463-1
58. Gugiatti, M., et al., "Towards the TRISTAN detector: Characterization of a 47-pixel monolithic SDD array", Nucl. Instrum. Meth. A 1025 (2022); doi.org/10.1016/j.nima.2021.166102
59. Max Planck Institute for Physics, "KATRIN and the TRISTAN detector", https://www.mpp.mpg.de/en/research/astroparticle-physics-and-cosmology/dark-matter-experiments/tristan-detector-search-for-sterile-neutrinos
60. Dolinski, M.J., et al., "Neutrinoless double-beta decay: status and prospects", Annual Review of Nuclear and Particle Science Vol. 69:219-251 (Volume publication date 19 October 2019); doi.org/10.1146/annurev-nucl-101918-023407
61. Rodejohann, W., "Neutrino-Less Double Beta Decay and Particle Physics", International Journal of Modern Physics E, Vol. 20, No. 09 (2011); doi.org/10.1142/S0218301311020186
62. Bilenky, S., "Neutrinos: Majorana or Dirac?", Universe Vol. 6 Iss. 9; doi.org/10.3390/universe6090134
63. The MAJORANA Collaboration, "The MAJORANA Neutrinoless Double-beta Decay Experiment", University of Washington; https://www.npl.washington.edu/majorana/majorana-experiment
64. Brofferio, C., et al., "Neutrinoless Double Beta Decay Experiments With $TeO_2$ Low-Temperature Detectors", Front. Phys., (2019); doi.org/10.3389/fphy.2019.00086
65. Armengaud, E., et al. (CUPID-Mo Collaboration), "New Limit for Neutrinoless Double-Beta Decay of 100Mo from the CUPID-Mo Experiment", Phys. Rev. Lett. 126, 181802 (2021); doi.org/10.1103/PhysRevLett.126.181802
66. Crane, L., "Physicists fail to find mysterious 'sterile neutrino' particles", NewScientist Physics (2021); https://www.newscientist.com/article/2294958-physicists-fail-to-find-mysterious-sterile-neutrino-particles/
67. Johnston, H., "Sterile neutrinos ruled out by MicroBooNE experiment", Phys. World 34 (12) 5i (2021); https://iopscience.iop.org/article/10.1088/2058-7058/34/12/05/pdf
68. Argüelles, C. A., et al., "MicroBooNE and the $\nu_e$ Interpretation of the MiniBooNE Low-Energy Excess", CERN-TH-2021-195, FERMILAB-PUB-21-618-T, IPPP/21/50, FTPI-MINN-21-23; doi.org/10.48550/arXiv.2111.10359
69. Aartsen, M. G., et al. (IceCube Collaboration), "eV-Scale Sterile Neutrino Search Using Eight Years of Atmospheric Muon Neutrino Data from the IceCube Neutrino Observatory", Phys. Rev. Lett. 125, 141801 (2020); doi.org/10.1103/PhysRevLett.125.141801
70. Aker, M., et al. (KATRIN Collaboration), "Bound on 3+1 Active-Sterile Neutrino Mixing from the First Four-Week Science Run of KATRIN", Phys. Rev. Lett. 126, 091803 (2021); doi.org/10.1103/PhysRevLett.126.091803